\documentclass[11pt]{iopart} 
\input epsf  
\begin{document}
\jl{1}

\title[The Active Baker Map]
{Autocatalytic Reactions in Systems with
Hyperbolic Mixing: Exact Results for the 
Active Baker Map}

\author{Z Toroczkai\dag, 
G K\'arolyi\ddag, \'A P\'entek\S, and
T T\'el$\|$}

\address{\dag\  Theoretical
Division and Center for Nonlinear Studies,
Los Alamos National Laboratory, Los Alamos, 
New Mexico 87545, USA}

\address{\ddag\ Department of Civil Engineering Mechanics,
Technical University of Budapest, M\H{u}egyetem rkp.~3,
H-1521 Budapest, Hungary}

\address{\S Marine Physical Laboratory, Scripps Institution
of Oceanography, University of California
at San Diego, La Jolla, CA 92093-0238, USA}

\address{$\|$\ Institute for Theoretical Physics,
 E\"{o}tv\"{o}s
University, P. O. Box 32, H-1518 Budapest, Hungary}


\ead{toro@lanl.gov, karolyi@tas.me.bme.hu, apentek@ucsd.edu, 
tel@poe.elte.hu} 

\begin{abstract}
We investigate the effects of  hyperbolic 
hydrodynamical mixing
on the reaction kinetics of autocatalytic systems.
Exact results are derived for the two dimensional open
baker map as an underlying mixing dynamics for a two-component
autocatalytic system, $A+B \to 2B$. We prove that the hyperboliticity
exponentially enhances the productivity of the reaction which is 
due the fact that the reaction kinetics is catalyzed by the 
fractal unstable manifold of the chaotic set 
of the reaction-free dynamics. 
The results are compared with phenomenological 
theories of active advection.
\end{abstract}

\submitto{\JPA}



\section{Introduction}
\label{sec:intro}

The advection of chemically or biologically active 
particles in 
imperfectly mixing chaotic hydrodynamical flows 
received considerable recent
interest \cite{ottino,prl,pre,physa,NLH,chaos,NLHT,Spall,
Bracco}. 
 By chaotic we understand a smoothly time-dependent, 
in the simplest case
periodic, flow in which the advection dynamics of
tracers is chaotic.
These tracers undergo chemical or biological interactions while
being advected by two-dimensional flows. The  distribution of
the products along a filamental fractal seems 
to be a general feature of such reactions.

Here we consider tracers as point-like particles 
which are
assumed not to have any influence
on the fluid flow itself. The reaction model is
of kinetic type where the reactions take place upon 
particle-particle `collisions':
when particles of type 
$A$ and $B$ are closer to each other than a 
\emph{reaction range} $\sigma$, then particle $A$
changes 
to $B$ 
according to the autocatalytic scheme $A+B \to 2B$.
The reaction model is discrete in time, i.e., a 
reaction can only 
take place at integer multiples of a \emph{reaction time}
$\tau$. In intervals between successive reactions the particles
are passively advected by the flow. 

The passive advection dynamics
has two essential parameters that influence 
the reaction kinetics of
the process: a decay or contraction parameter 
$\lambda$
and the fractal dimension $D_0$ of the set around which 
products accumulate.
In open flows  
this set turns out 
to be the unstable manifold of a
nonescaping chaotic set characterizing the passive advection
\cite{prl,pre}, and
the role of the contraction parameter 
is played by the average negative Lyapunov  exponent of the 
Lagrangian dynamics which expresses the convergence towards
the chaotic set. Due to incompressibility, the positive and negative
Lyapunov exponents have the same absolute value.
Particles are advected away from any neighborhood of the
chaotic set along the unstable manifold which leads 
to an exponential decay of
the passive particles with an escape rate $\kappa$.
This is, however, not 
a new dynamical parameter 
since it is related to the previous ones 
via $D_0 \approx 2-\kappa/\lambda$,
\cite{T0,tranchaos}. Here we concentrate on \emph{slow} 
reactions whose reaction velocity {\bf $v_r=\sigma/\tau$}
is much slower than the characteristic velocity of the
passive advection dynamics.

\begin{figure}[htbp] 
\hspace*{1.3cm}
\begin{minipage}{5.0 in}\epsfxsize=5.0 in
\epsfbox{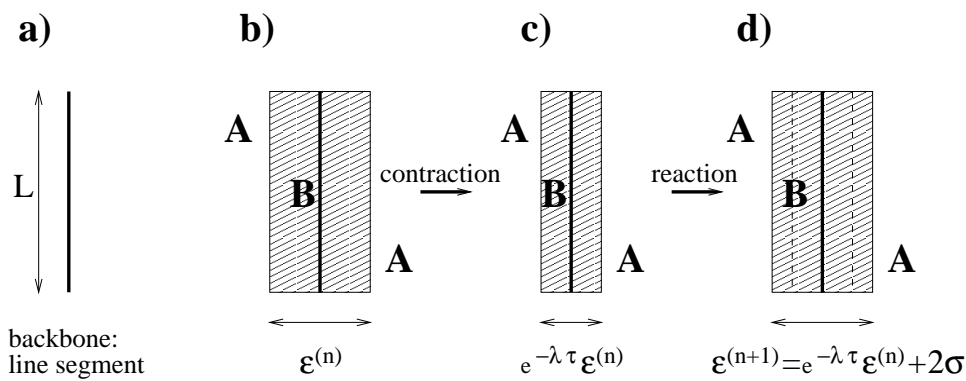}
\end{minipage}
\vspace*{0.5cm} 
\caption{ Schematic drawing of the emptying and fattening
process surrounding a single filament. A line segment of length $L$
( a)) is covered by material $B$ ( b)). The width of this stripe is
first shrunk ( c)) due to outflow, then it is fattened by a sudden reaction
( d)).}\label{fig:sema} 
\end{figure}

In this section we present 
a heuristic argument 
for  the total area of
the product particles as a function of time. 
Since the basic ingredients of the argument  are
the existence of a fractal backbone and the 
contracting dynamics acting towards the backbone, we
do not need to actually specify the  two-dimensional hydrodynamical
flow itself just to postulate the fractal backbone 
and the contraction property.
 We consider the simplest filamental fractal which is the direct 
product of a Cantor set of dimension 
$(D_0-1)$ and a line segment of length $L$
so that
the total dimension is $D_0$. 
In addition, a contraction  of
rate $\lambda$ is postulated 
towards the fractal 
filaments, see Figure~\ref{fig:sema}. 
This simple theory can also be considered to hold
for reactions taking place along spatially fixed fractal catalysts
when there is  transport of  products, resulting
in an effective contraction.

At time $n$, right after a reaction takes place,
filaments are assumed to be covered by material $B$
in stripes of average width $\varepsilon^{(n)}$. 
The complement of these strips is
assumed to be covered by material $A$ everywhere.
During the next period of length $\tau$ no 
reaction takes place, only
contraction, which reduces the
widths $\varepsilon^{(n)}$ by a factor of
$\exp{(-\lambda \tau)}$.  
The effect of the reaction is simply an `infection'
by $B$ particles of the $A$ neighbors
in circles of radii $\sigma$ which in case of a filamental
geometry leads to the 
broadening of the stripe width by {$g \sigma$, $g = 2$}.
Thus we obtain a recursion relation 
expressing the stripe widths 
$\varepsilon^{(n+1)}$ right after
the $(n+1)$-st reaction in terms of $\varepsilon^{(n)}$, 
right after the $n$-th reaction: 
\begin{equation}
\varepsilon^{(n+1)} = e^{-\lambda \tau}
\varepsilon^{(n)} +g\sigma .
\label{rec}
\end{equation}
This simple recursive dynamics possesses a 
single fixed point attractor
\begin{equation}
{\varepsilon^*} =
\frac{g \sigma}{1-e^{-\lambda \tau}}
\label{eq:MB0}
\end{equation}
which is reached for  $n \rightarrow \infty$.
This shows that the product distribution reaches saturation:
an active steady state sets in after some time in which 
the activity exactly cancels the material 
loss due to contraction.

Although there are infinitely many filaments 
in the fractal backbone,		
if they are  covered with stripes of non-zero  width,
the number of such covered filaments becomes
finite due to the overlaps among them.
It follows from the properties of fractals 
that the number $N(\epsilon)$ of filaments  covered
by width $\epsilon$ is
\begin{equation}
N(\epsilon) \sim \left( \frac{\epsilon}{L} \right)^{-(D_0-1)}.
\end{equation}
The total area $\cal{ A}$ of these covered filaments is then
\begin{equation}
{\cal A}(\epsilon)= \epsilon L N(\epsilon)
\sim {\epsilon}^{2-D_0} L^{D_0}. \label{aaa}
\end{equation}
Next we apply this result with $\epsilon=
\varepsilon^{(n)}$ at time
$n$ and with $\epsilon=\varepsilon^{(n+1)}$ at time $n+1$.
Note that we are then working on the egde of 
the validity of fractal scaling
since below the scale of $\varepsilon^{(n)}$ or 
$\varepsilon^{(n+1)}$
the product distribution is two-dimensional. 
The consequences of
the crudeness of this approximation will  be analyzed later,
in view of the exact results
obtained for the baker map.
Using (\ref{aaa}),  
one finds a recursion for the area ${\cal A}_B$
of $B$ particles. It follows from (\ref{rec}) that 
\begin{equation}
{\cal A}_{B}^{(n+1)}= 
  \left\{ e^{-\lambda \tau}\left[{\cal A}_{B}^{(n)}\right]^{1/(2-D_0)} +
  g \sigma L^\frac{D_0}{2-D_0} \right\}^{2-D_0}.
  \label{eq:mapauto}
\end{equation}
In  the case of exactly parallel filaments 
the factor $g$ is identically 
$g=2$ before the stripes start to overlap. 
After overlap takes place,
which is unavoidable on sufficiently fine scales, the value of
$g$ may however differ from 2. 
In this approach, we assume that the 
overlap effects can be taken care by introducing an \emph{average}
geometrical factor $g$, which is \emph{independent} of the
reaction parameters $\sigma$ and $\tau$.
If the proportionality factor in
(\ref{aaa}), which is the
Hausdorff measure of the fractal set, is different from 1,
it may also be incorporated in the
geometrical factor $g$.
The form of  recursion (\ref{eq:mapauto}) is a lot more complicated 
than recursion (\ref{rec}) for the
stripe widths due to the fractal character of 
the backbone. It shows that the `microscopic' width 
dynamics and the `macroscopic' area dynamics are 
qualitatively different 
since the latter is nonlinear. Nevertheless, 
(\ref{eq:mapauto}) possesses a fixed point
\begin{equation}
{\cal A}_{B}^{*}= 
  \left(\frac{g \sigma/L}{1-e^{-\lambda \tau}} \right)^{2-D_0} L^2
  \label{eq:map*}
\end{equation}
which is the only attractor of the chemical dynamics. 

It is worth taking a continuum limit
in which the reaction time $\tau$ and distance 
$\sigma$
both  go to zero, while
the reaction velocity $v_r \equiv \sigma/\tau$
remains finite. In this limit a differential 
equation is obtained for the $B$ 
area ${ \cal A}_{B}(t)$ as a 
function of the continuous time $t$ in the form
\begin{equation}
\dot{{\cal A}}_B=-\lambda (2-D_0) {\cal A}_B+
g (2-D_0) v_r L^\frac{D_0}{2-D_0}
  \left(
{\cal A}_B
\right)
^{-\beta} \label{diffeq}
\end{equation}
where the exponent $\beta$ is a unique positive expression of the
fractal dimension
\begin{equation}
\beta =\frac{D_0-1}{2-D_0} \label{beta}.
\end{equation}
The singular-looking reaction  term on 
the right hand side (rhs) of (\ref{diffeq}) expresses the fact that
the smaller the area, the larger the production rate becomes,
just as one expects for reactions on a fractal catalyst. 

Both the discrete-time  and the continuous-time area dynamics possess 
a fixed point steady state. The corresponding product area of the
latter is
\begin{equation}
{\cal A}_B^* = \left( 
\frac{g v_r}{\lambda L} \right)^{(2-D_0)} L^2.
\label{contarea}
\end{equation}
Note that this is  much larger than the product area along a single
non-fractal line (obtained as the special case $D_0=1$)
since the expression in the paranthesis is much 
less than unity in view
of our assumption of the slowness of the active dynamics.
This clearly indicates that the presence of a fractal 
catalyst makes the reactions much more efficient. 
The important caveat is that
the fractal backbone is not introduced `by hand' 
as an imposed mixing rule, it
is rather generated naturally by the 
\emph{chaotic advection dynamics} of 
the fluid flow transporting the tracers.
The chaotic set 
is the union of unstable bounded orbits. The reactions take
place along
the outflow curves from these orbits (the unstable manifold).
For a time periodic flow
the unstable manifold moves periodically in the plane of the
flow (but does not change its fractal dimension). 
This is, however, not a basic difference since on a stroboscopic map 
the manifold appears to be still.

In order to illustrate and also check the 
above general heuristic
arguments, 
in this paper we consider two variants 
of the baker map, which describe
a discrete mixing dynamics with hyperbolic 
chaotic properties. These maps
are simple enough to facilitate an exact 
analytic approach and yet
they incorporate the generic features of mixing 
from realistic flows. Within this model we are able
to derive exact expressions for the area dynamics and thus to identify
the geometric factor, which was not possible
within the phenomenological theories of active chaos.	 

Our paper is organized as follows: in Section \ref{sec:active}
we introduce a simple variant of the baker map on which we 
superimpose the chemical activity, introducing thus a novel
reactive dynamical system, the \emph{Active baker map}.
Then we derive the exact stripe dynamics for this map, including
the reaction equation describing the area dynamics of the 
reactants in the system. In Section \ref{sec:compare}
we compare the exact expressions with the heuristic results
given in the Introduction, identifying the exact effects 
of the fractal geometry on the geometric factor. In Section
{\ref{sec:another}} we consider a second variant for
the baker map, and show the robustness of the heuristic
results. Section \ref{sec:discussion} is devoted to discussions
and outlook.


\section{The active baker map}
\label{sec:active}

The passive (reaction-free) baker map 
is probably the simplest dynamical system which exhibits
chaotic behaviour of purely hyperbolic type.
Its simplicity allows for analytic manipulations,
and thus it plays from a theoretical standpoint
a crucial role, comparable to the Ising model from
statistical mechanics.  There are several variants to this map, and
we shall consider two of them in more detail.  
Let us give first one variant of the baker map and then define the
active version based on this, then in Sec.~\ref{sec:another} introduce
another variant with the associated active counterpart, and then
finally compare the results based on the actual 
realizations of the maps.

Assuming
that we start from a completely filled unit square, one step of the
baker map consists of the following:
1) the lower (upper)
half of the unit square is compressed in the $x$ direction
by a factor of $a < 1/2$ around
the fixed point in $(0,0)$ ($(1,1)$), and then 2) the compressed
stripes are
stretched along the $y$ direction (around the same fixed points)
by a factor of $b \geq 2$.
For $b>2$ there will be
`material' reaching over the unit square, 
which is clipped and discarded. In this case the map is modeling an open flow.
By taking the unit square as the phase space of the baker map,
the characteristic length $L$ of the dynamics is chosen to be unity.
When $ab = 1$, the baker map is area preserving
in the sense that its Jacobian is uniformly unity, however, it is
an open chaotic map for any $b > 2$, the chaos being of 
transient type \cite{tranchaos}.
Mathematically, the image $(x',y')$ through the baker map
 of a point $(x,y)$
from the unit square is given by:
\begin{eqnarray}
\begin{array}{l}
x'=ax+(1-a)\theta\left(y-1/2\right)\;,\;\;\;\;x\in[0,1] \\
y'=by-(b-1)\theta\left(y-1/2\right)\;,\;\;
\;\;\;y\in[0,1] \label{bakerorig}
\end{array} 
\end{eqnarray}
where $\theta(x)$ is the Heaviside step function.
The map is not defined outside the unit square (i.e.\ $L=1$)
and thus chemical reactions can only occur inside the unit square.
In the following we restrict
ourselves to open area preserving systems, i.e., to the regime $b=1/a>2$.
Repeated application of the map will result in a sequence of vertical 
stripes 
of various widths which form the hierarchy
of a one dimensional Cantor set in the unit interval along the $x$-axis.
Contraction towards the vertical filaments is governed by the
Lyapunov exponent
\begin{equation}
\lambda =-\ln a>0\;.     \label{lambda}
\end{equation}
Since due to the construction, in the vertical direction there is
no structure, 
the map is effectively one 
dimensional.
The fractal dimension $D_0-1$ of the Cantor set
on which the stripes reside, is simply calculated to be:
\begin{equation}
D_0-1=-\frac{\ln{2}}{\ln{a}}\;.
\label{dimension}
\end{equation}
The decay of the area covered by the  material left in 
the unit square is exponential,
with a rate given by the so called escape rate, $\kappa$:
\begin{equation}
\kappa = - \ln{(2a)}\;, \label{kappa}
\end{equation}
so that $D_0=2-\kappa/\lambda$ holds for the fractal dimension
of the stripes in the unit square.
Initially, the empty space between the stripes
is imagined to be filled with `material'
$A$, while 
the stripes contain material $B$.

We call the baker map \emph{active}
when the two materials react with
each other. We consider autocatalytic reactions $A+B \to 2B$
(the same as in the Introduction), with
the reactions happening along the (vertical)
contact lines. This reaction can alternatively
be considered simply as the surface-growth of 
material $B$, and leads to a 
broadening of the stripes relative to those of the passive baker map.

We assume that
the reactions happen instantaneously and 
simultaneously in the system, \emph{after 
every} $\tau$ steps of the reaction-free baker map.
Time is measured in units of the baker-steps. 
Thus, for $\tau=1$,
the $B$ surface grows after each baker-step by an amount of
$\sigma$ on each side of a stripe. For $\tau=2$, two 
reaction-free baker-steps are
taken before one instantaneous reaction occurs, etc. 

\begin{figure}[htbp] 
\hspace*{-0.1cm}
\begin{minipage}{5.8 in}\epsfxsize=5.8 in
\epsfbox{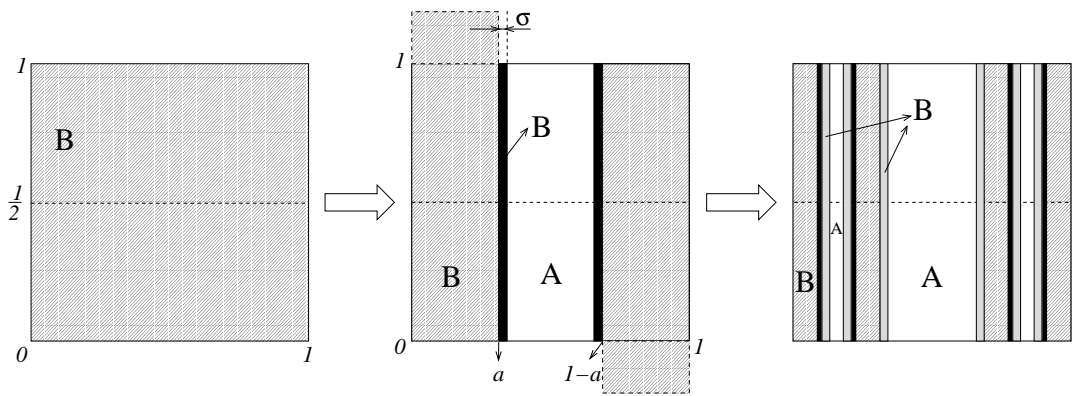}
\end{minipage}
\vspace*{0.5cm} 
\caption{Two consecutive steps of the baker map and two reactions
($\tau=1$).
The bands of width $\sigma$ become occupied by $B$ in each reaction.   
The material hanging over the unit square is discarded.
}\label{fig:themap} 
\end{figure}

For the simplicity of writing, 
let us introduce the following 
convention: passive steps will be called
\emph{baker-steps}, and the combination of $\tau$ baker-steps
followed immediately by one reaction will be 
called one \emph{active-step}.

If the gap between two stripes is shorter than $2\sigma$,
the gap disappears
during reaction, and the stripes merge into a single one.
In the absence of reactions ($\sigma=0$) the steady state (defined as
the asymptotic state $t \to \infty$) is characterized by a
zero Lebesque-measure distribution of material $B$,
covering the \emph{unstable manifold of teh chaotic set}
of dimension $D_0$ of the baker 
map. 
In the case of the active baker map, however,
there are new $B$-areas of width
$\sigma$ appearing after periods $\tau$ in the system, and thus
a sort of dynamical balance is expected to set in
between the escape dynamics and the growth dynamics.


\subsection{Stripe hierarchy in a simple case: 
one baker-step per active-step}

In spite of the fact that the baker map is one of the
simplest maps that exhibits chaotic dynamics, when
chemical activity is also introduced, the reaction
kinetics of the system becomes rather involved.

Let us first start with the
simpler case of one reaction per baker-step, i.e.,
$\tau = 1$. 
Figure~\ref{fig:themap} shows the succession of two such active-steps.
The black bands show the $B$ material coming from the first reaction,
while the dotted light gray bands are the  $B$ material 
emerging  from the
second reaction. {From} the point of view of the reaction kinetics,
the distribution and the amount of material $A$ (the gaps)
is crucial. If an $A$ stripe is less wide than
$2\sigma$ then it will fill up in the next reaction and disappears.
During the baker-steps the hierarchy of gaps builds up with
thinner and thinner gaps appearing. {From} a certain level on,
the reactions are putting a stop to the hierarchy growth.
The question is how in the long time limit the picture
stabilizes and the system settles into a stationary
state.
Every gap is bounded from left and right by material $B$.
Since the contraction by $a$ is a point transformation
(around the two fixed points at the edges),
each gap will shrink by a factor of $a$ and it will also get
closer to the corresponding fixed point by the same factor
$a$. After the reaction the gaps will lose $2\sigma$ from their
width.
After one ($n=1$) active-step we get one gap
of width $1-2a-2\sigma$; after two active-steps we obtain
three gaps: one of width $1-2a-2\sigma$, and two gaps, each
of width $a(1-2a-2\sigma)-2\sigma$; for $n=3$, we have a total
of 7 gaps, three just like in the $n=2$ case and 4 new members
each of width $a[a(1-2a-2\sigma)-2\sigma]-2\sigma$, etc. 
An important observation is that if gaps appear at a certain
level of the hierarchy, there always will be gaps of precisely the
same width and position in the unit square at any
later time:
during the next active-step, a particular gap
will \emph{map} precisely down into the location of two other gaps
of the given level. 
The largest gap (the top level of the hierarchy)
is always located in between the points $x_l = a+\sigma$ and
$x_r = 1-a-\sigma$. During the next active-step this particular
gap is mapped into two lower level gaps, however its original
position and width is taken over by the \emph{new} gap appearing
as the images of the edges at $x=0$ and $x=1$ of the unit square,
and modified by $\sigma$ due to the reaction.

Thus the hierarchy being formed can be described as follows.
After $n$ active-steps, there will be gaps of type $n$ (i.e., the
hierarchy will have $n$ levels).
If $w_1^{(j)}$ denotes the width of a gap on level $j$,
and $m_1^{(j)}$ denotes their number (or multiplicity), then one can
write:
\begin{equation}
w_1^{(j)} = (1-2a)a^{j-1}-2\sigma \frac{1-a^j}{1-a}
\;,\;\;\;\; m_1^{(j)}=2^{j-1}\;. \label{gaps}
\end{equation}
Observe that the first term on the rhs of the first equation
is exactly the width of the gaps in the reaction free baker
map ($\sigma=0$). 
The second represents the total amount 
of area-decrease of each gap in $j$ reaction steps:
$-2\sigma(1+a+a^2+...a^{j-1})$.
The total area occupied by the gaps (or material
$A$) after $n$ active-steps is given by
\begin{equation}
{\cal A}_A^{(n)}=\sum_{j=1}^{n}m_1^{(j)} w_1^{(j)}=
1-(2a)^n - \left(
2^n-1 - a \frac{1-(2a)^n}{1-2a}
\right) \frac{2\sigma}{1-a}\;.  \label{gaparea}
\end{equation}
The area occupied by material $B$ is simply the
complementer of the above in the unit square:
\begin{equation}
{\cal A}_B^{(n)} = 1-{\cal A}_A^{(n)}=
(2a)^n + \left(
2^n-1 - a \frac{1-(2a)^n}{1-2a}
\right) \frac{2\sigma}{1-a}\;.  \label{Barea}
\end{equation}
When $n \to \infty$, ${\cal A}_A^{(n)}$ becomes negative, and
later both areas formally diverge which 
indicates that this process has to stop
after a finite number of steps.
This happens by the closing of the smallest gaps
due to the reactions in the $n^*$-th step:
no further gaps can be created from a certain level down
in the hierarchy because the reactions are of non-zero, finite 
width $\sigma$.
By definition, the thinnest gaps of an $n$-level hierarchy are 
the gaps at the 
lowest ($n$-th) level of width
$w_1^{(n)}$.
The condition for the lowest level 
to be filled in $n^*$ active-steps
is $w_1^{(n^*-1)} > 0$ and $w_1^{(n^*)}\leq 0$.
{From} this, and Eq. \ref{gaps},
we obtain
\begin{eqnarray}
n^* = \left\{
\begin{array}{ll}
[u]+1\;,   & \mbox{if}\;\; u
\;\;\mbox{is non-integer}\;,\\
u\;, & \mbox{if}\;\; u
\;\;\mbox{is an integer}\;,
\end{array} \right. \label{tt1}
\end{eqnarray}
with
\begin{equation}
u=\frac{\ln{\left(1+\frac{(1-a)(1-2a)}
{2\sigma a} \right)}}
{\ln{\frac{1}{a}}}\;,  \label{thinnest}
\end{equation}
($[x]$ denotes the integer part of $x$).
For $n \geq n^*$, $w_1^{(n)} \leq 0$ and there are no more
gaps at and below this level. Thus, even if the active-steps
are repeated indefinitely, the hierarchy of gaps
stops at $j=n^*$. Since the gaps at higher
level map into gaps at lower level, a steady state
is reached after $n^*$ steps, and the whole 
material distribution becomes
stationary. The fact that a stationary state is reached as 
a balance between the emptying dynamics and the reactions
was numerically observed in all the previously studied dynamical
systems (e.g., flow in the wake of a cylindrical obstacle) 
\cite{prl,pre}, corroborating the generic character of the
stationary state.

\subsection{Stripe-hierarchy for multiple 
baker-steps per active-step}

In this case there are $\tau$ baker-steps
taken before an instantaneous reaction occurs. Let us describe
the hierarchy of gaps built up during
$n$ active-steps. After the first $\tau$ baker-steps we obtain
the usual reaction-free baker hierarchy with
$\tau$ different types of gaps (a total of $2^{\tau}-1$ gaps).
After the first reaction there will be therefore
$m_k^{(1)}=2^{k-1}$ gaps of type $k$ whose width is
$w_k^{(1)} = a^{k-1}(1-2a)-2\sigma$ 
with $k=1,2,..,\tau$. 
Following the next active-step,
one obtains $2\tau$ types of gaps.
The first $\tau$ types of gaps are generated by
the $k$ times iterated images of the endpoints
$x=0$ and $x=1$ ($k=1,2,..,\tau$) and modified by the reaction.
These have multiplicities
widths and positions exactly as those after the first active-step.
The second
$\tau$ types of gaps are $\tau$ times iterated images
of the gaps coming from the first active-step.
The widths and multiplicities of this second set are given by
$w_k^{(2)} = a^{\tau}w_k^{(1)}-2\sigma$, and
$m_k^{(2)}=2^{\tau+k-1}$ on each level $k=1,..,\tau$.
In general, if we perform $n$ active-steps, there will be a total
of $2^{n\tau}-1$ gaps generated with the following widths and 
multiplicities:
\begin{eqnarray}
\begin{array}{ll} w_k^{(1)} = a^{k-1} (1-2a)-2\sigma\;, &
m_k^{(1)}=2^{(k-1)}\;,\\
w_k^{(j)} = a^{\tau}w_k^{(j-1)}-2\sigma\;, &
m_k^{(j)}=2^{(j-1)\tau+k-1}\;,
\end{array} 
\label{taurecursion}
\end{eqnarray}
where $k=1,..,\tau$ and $j=2,..,n$.
Analogously to the
$\tau=1$ case, during the iteration,
the gap-hierarchy changes by the 
\emph{addition} of finer gaps, and the previously created
gap-structure is \emph{not modified} by the new reaction.
However, in contrast to the $\tau=1$ case the new gaps
resulted from the new reaction form a \emph{sub-hierarchy}
of $\tau$ levels. In the expressions (\ref{taurecursion})
$j$ denotes the main levels of the hierarchy (related to
the state just after a reaction) while $k$ quantifies
the hierarchy on the subtree level (created by the
in-between-reactions baker-steps).
Recursion (\ref{taurecursion}) is solved easily, and the result is:
\begin{equation}
\left.
\begin{array}{lll}
w_k^{(j)} & = & (1-2a) a^{(j-1)\tau+k-1}
-2\sigma \frac{1-a^{j\tau}}{1-a^\tau}\;,\\ 
m_k^{(j)} & = & 2^{(j-1)\tau+k-1}\;, 
\end{array}
\right\}
j=1,...,n\;.\label{tausol}
\end{equation}
Obviously the widths are decreasing functions
of both $j$ and $k$ (down the hierarchy) as also seen
from (\ref{tausol}). 
Therefore, after 
$n$ active-steps the thinnest
gaps will have the width of $w_{\tau}^{(n)} = 
(1-2a)a^{n\tau-1}-2\sigma (1-a^{n\tau})/(1-a^{\tau})$
and multiplicity of $m_{\tau}^{(n)} = 2^{n\tau -1}$.
Another important consequence of the monotonic
behaviour is that the
widths found within a level $j$ are \emph{all} smaller
than the widths found within the level $j-1$.
This statement is easily proven by comparing
the thinnest and thickest gaps of the neighboring levels.

Thus, the total area of material $A$ after $n$ active-steps
is computed as:
\begin{equation}
\fl {\cal A}_A^{(n)} = \sum_{j=1}^{n} \sum_{k=1}^{\tau}
m_k^{(j)} w_k^{(j)} =
1-(2a)^{n\tau} - \left(
2^{n\tau}-1 - a^{\tau}\left( 2^{\tau} -1\right)
\frac{1-(2a)^{n\tau}}{1-(2a)^{\tau}}
\right) \frac{2\sigma}{1-a^{\tau}}\;. \label{Atau}
\end{equation}
For $\tau=1$ we indeed recover Eq.~(\ref{gaparea}).
The total amount of material
$B$ left in the system will again be the complementary
of (\ref{Atau}) in the unit square, i.e., 
${\cal A}_B^{(n)} = 1-{\cal A}_A^{(n)}$:
\begin{equation}
{\cal A}_B^{(n)} = (2a)^{n\tau} + \left( 
2^{n\tau}-1 - a^{\tau}\left( 2^{\tau} -1\right)
\frac{1-(2a)^{n\tau}}{1-(2a)^{\tau}}
\right) \frac{2\sigma}{1-a^{\tau}}\;. \label{Btau}
\end{equation}
As $n$ (time) increases, there will be
a critical $n^* = n^*(\tau)$ value at which gaps start
to disappear for the first time. 
Of course, Eq.~(\ref{Btau}) is valid only in the regime where
no gaps have been filled yet. 
This means that for 
$n = n^*-1$ all the gaps of the hierarchy
(\ref{tausol}) exist, but for $n=n^*$ \emph{some} of the
gaps have disappeared. 
Let us calculate this value
and see how a stationary state is reached.
Since the thinnest gaps of width $w_{\tau}^{(n)}$
must have disappeared at $n=n^*$ (i.e., 
$w_{\tau}^{(n^*)} \leq 0$), 
but they were
present at  $n = n^*-1$ (i.e., $w_{\tau}^{(n^*-1)}>0$),
$n^*$ is calculated as
\begin{eqnarray}
n^* = \left\{
\begin{array}{ll}
[u]+1\;,   & \mbox{if}\;\; u
\;\;\mbox{is non-integer}\;,\\
u\;, & \mbox{if}\;\; u
\;\;\mbox{is an integer}\;,
\end{array} \right. \label{tt}
\end{eqnarray}
where
\begin{equation}
u=\frac{\ln{\left(1+\frac{(1-a^{\tau})(1-2a)}
{2\sigma a} \right)}}
{\ln{\frac{1}{a^{\tau}}}}\;.  \label{tstart}
\end{equation}
This gives the start of the filling
in terms of the number of active-steps. However, 
it is not 
necessary that only the thinnest gaps of $n^*$
have been filled. 
In principle,
all the gaps of level $n^*$ can be filled because, 
as observed above,
the widths of gaps at a higher level
are \emph{all} larger than those of a lower level.
How many gaps of level $n^*$ are filled does not follow from 
(\ref{tstart}).

Performing another active-step, we may still obtain
a partially filled sub-hierarchy on the main level
$n^*+1$, and similarly in the next step, etc. 
However, since the widths of the gap additions
is strongly decreasing, this process will eventually stop.
Once stopped, the whole material distribution becomes stationary
and the system is settled into a steady state.

The system reaches the steady state when also the thickest 
($k=1$) gaps are filled on the next main level.
This happens at $m^*$ for which
$w_1^{(n)} > 0$ if $n < m^*$, but $w_1^{(m^*)} \leq 0$.
This leads to:
\begin{eqnarray} 
m^* = \left\{ 
\begin{array}{ll} 
[s]+1\;,\;\;\; & \mbox{if}\;\; s
\;\;\mbox{is non-integer}\;,\\ 
s\;,\;\;\;  & \mbox{if}\;\; s
\;\;\mbox{is an integer}\;, 
\end{array} \right. 
\end{eqnarray} 
where
\begin{equation} 
s=\frac{\ln{\left(1+ \frac{1}{a^{\tau}}\frac{(1-a^{\tau})(1-2a)} 
{2\sigma} \right)}} 
{\ln{\frac{1}{a^{\tau}}}}\;.  \label{end} 
\end{equation} 

Next we compare $m^*$ and $n^*$ in order to see
how many steps are needed to reach the steady state, once a
filling begins at $n^*$.
{From} Eqs.~(\ref{tstart}) and (\ref{end}) it is obvious
that $s > u$
(the numerator on the rhs of (\ref{end}) is larger than that of 
(\ref{tstart}) as $\tau > 1$). 
Let us introduce the notation:
\begin{equation}
v \equiv \frac{(1-a^{\tau})(1-2a)}{2\sigma
}\;. \label{ve}
\end{equation}
The quantities $s$ and $u$ then become:
\begin{equation}
s = 1+
\frac{\ln{\left(a^{\tau}+v\right)}}
{\tau \ln{\left(\frac{1}{a}\right)}}\;,\;\;\;\;
\mbox{and}\;\;\;\;
u = \frac{1}{\tau}+
\frac{\ln{\left( a+v \right)}}
{\tau \ln{\left(\frac{1}{a}\right)}}.
\end{equation}
Simple manipulations lead to:
\begin{equation}
s -  u = 1-
\frac{1}{\tau} \left( 1+
\frac{\ln{ \left( \frac{a+v}{a^{\tau}+v}
\right) }}
{\ln{\left(\frac{1}{a}\right)}}
 \right) < 1 \label{ineq}
\end{equation}
This means that we have the bounds 
$u < s < u+1$, when $\tau\geq 2$.
(For $\tau =1$, $s$ and $u$ are identical.)
This implies that
\begin{equation}
m^* \in \left\{ n^*, n^*+1
\right\}\;,
\label{diffone}
\end{equation}
i.e., the 
steady state is reached either
on level $n^*$,
or on the next level,  $n^*+1$.

\begin{figure}[htbp]
\hspace*{0.2cm}
\begin{minipage}{2.9 in}\epsfxsize=2.9 in  
\epsfbox{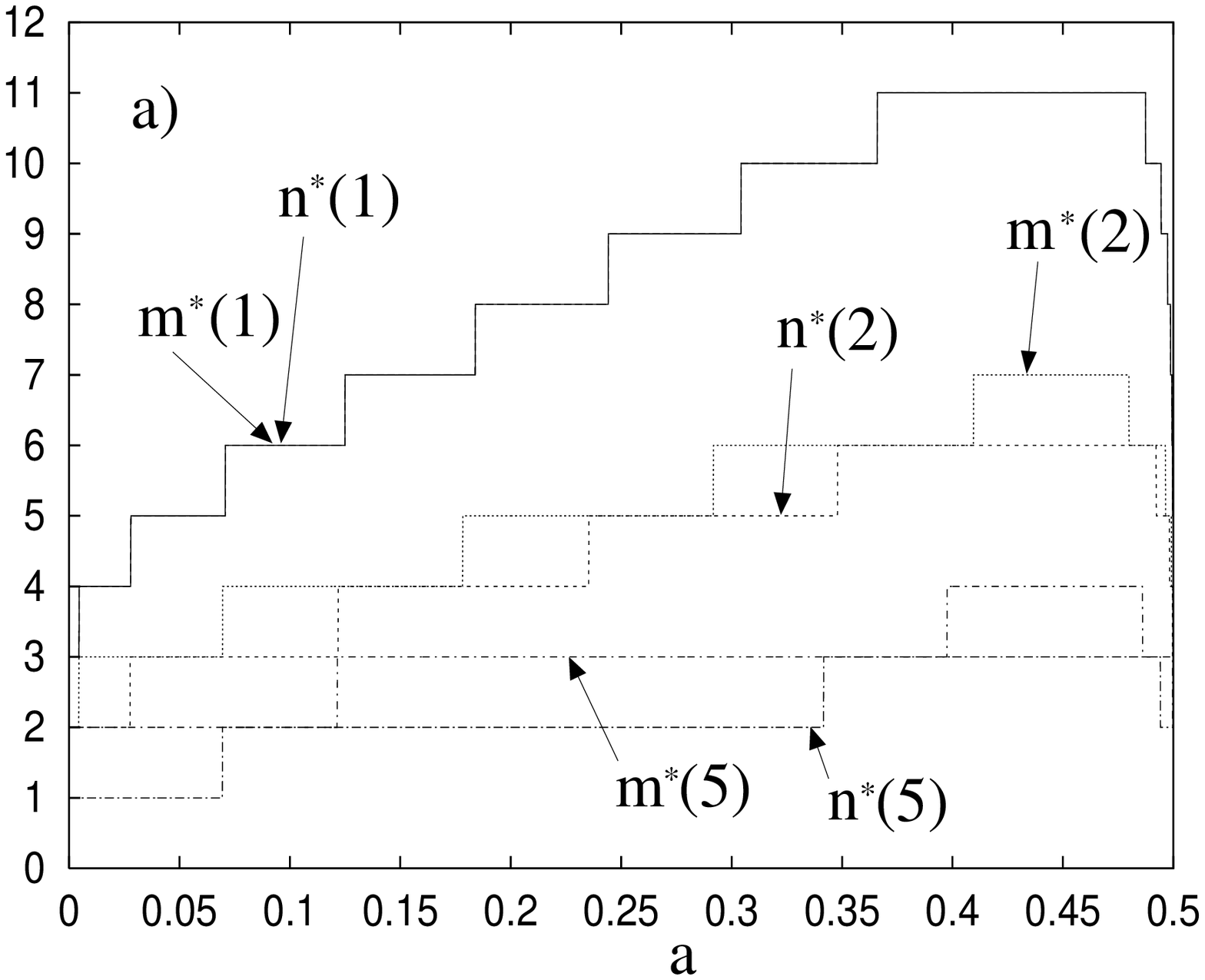} 
\end{minipage}
\begin{minipage}{2.9 in}\epsfxsize=2.9 in  
\epsfbox{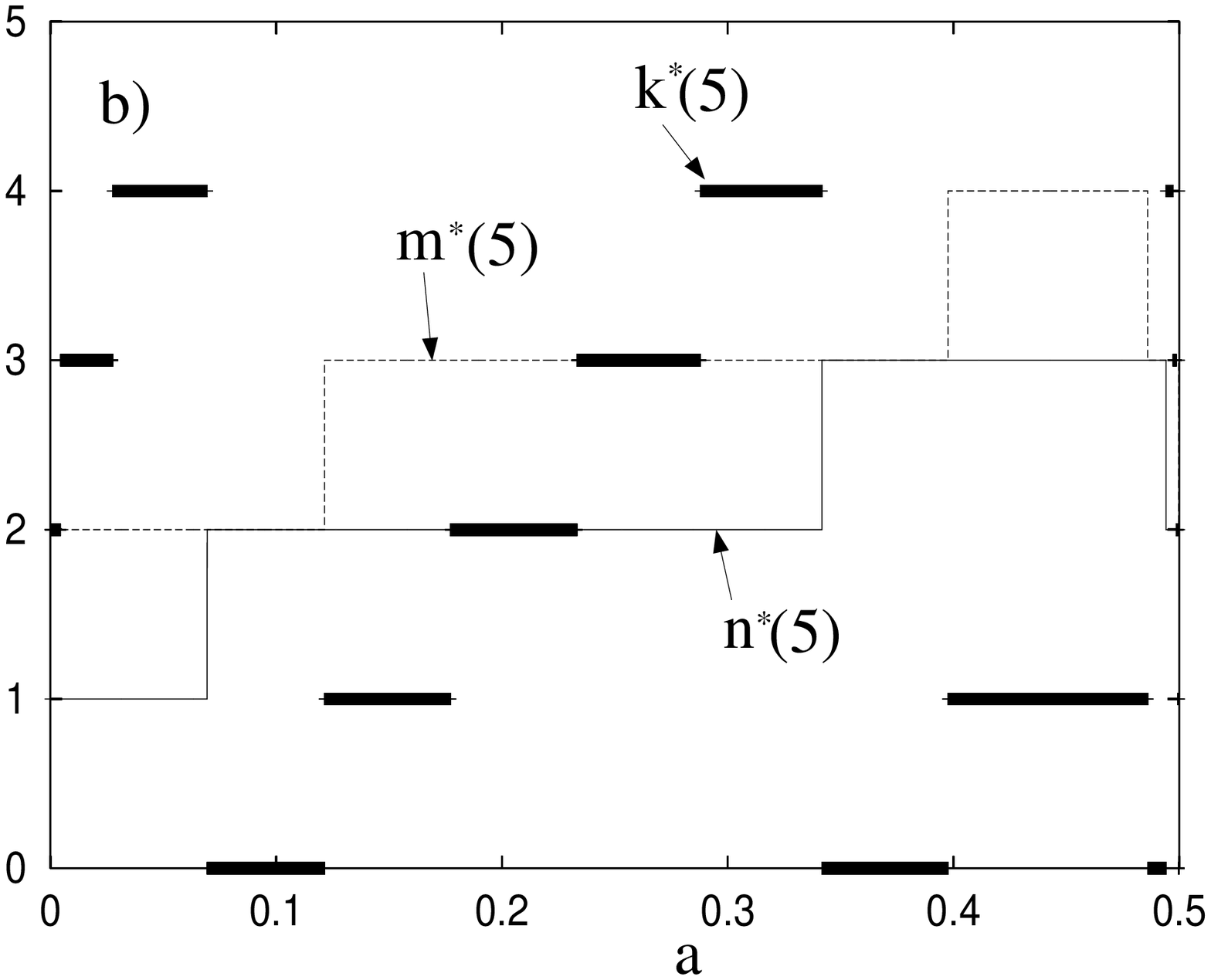} 
\end{minipage} 
\vspace*{0.0cm} 
\caption{a) Dependence  of $m^{*}$ and
$n^*$ on $a$ for three $\tau$ values: $\tau=1,2,5$.
Notice that relation (\ref{diffone}) holds. 
b) Dependence  of  $m^{*}$ and $k^*$ on $a$
for $\tau=5$. $k^*$ (diamonds) can take on any value
from the set $\{0,1,2,..,\tau-1 \}$. In all  
cases $\sigma=10^{-5}$. 
}\label{fig:mns} 
\end{figure} 

When $m^*=n^*$,
Eq.~(\ref{Atau}) correctly describes the time evolution
of  material $A$ over the whole range $n=1,2,..,n^*-1$,
since there are no partially filled main levels. Because there
are no new gaps added on the $n^*$-th step, in this case
${\cal A}_A^{(n^*)} = {\cal A}_A^{(n^*-1)}$ is the
steady state gap-area.

When $m^*=n^*+1$, the last main level is partially
filled, and in this case
${\cal A}_A^{(n^*+1)} = {\cal A}_A^{(n^*)}$
(on the $(n^*+1)$-st
step no new gaps are created)
is the steady-state gap-area:
\begin{equation}
{\cal A}_A^{(n^*)} = {\cal A}_A^{(n^*-1)}
+ \sum_{k=1}^{k^*} m_{k}^{(n^*)} 
w_{k}^{(n^*)} 
\label{extra}
\end{equation}
where $k^*$ is defined as the sub-level below which no gaps
exist within the main level $n^*$, i.e., $w_k^{(n^*)} > 0$
for $k \leq k^*$, and $w_k^{(n^*)} \leq 0$ for 
$k > k^*$. This results in:
\begin{eqnarray} 
k^* = \left\{ 
\begin{array}{ll} 
[y]+1\;,\;\;\; & \mbox{if}\;\; y
\;\;\mbox{is non-integer}\;,\\ 
y\;,\;\;\;  & \mbox{if}\;\; y
\;\;\mbox{is an integer}\;, 
\end{array} \right.\label{kval}
\end{eqnarray} 
where
\begin{equation} 
y=
\frac{\ln{\left( \frac{1-2a}{2\sigma }
\frac{1-a^{\tau}}{a^{\tau}}
\frac{a^{n^*\tau}}{1-a^{n^*\tau}}
\right)}}
{\ln{\left( \frac{1}{a}\right)}}\;. \label{rr} 
\end{equation} 
For the $m^*=n^*$ case we define $k^*$ to be zero.
Figure~\ref{fig:mns} shows the values of $m^*$, 
$n^*$
and $k^*$ as  function of
$a$ for a few particular $\tau$ values.

\subsection{Steady-state areas}

Next we analyze the steady-state expression of the
area occupied by material $B$.
When $m^*=n^*$, the steady-state value
of the $B$-area is given by ${\cal A}^{(n^*-1)}_B$,
and for $m^*=n^*+1$ by 
${\cal A}^{(n^*)}_B$, the complementary
of (\ref{extra}) in the unit square, i.e.,  
${\cal A}_B^{(n^*)} = {\cal A}_B^{(n^*-1)} 
- \sum_{k=1}^{k^*} m_{k}^{(n^*)} 
w_{k}^{(n^*)}$.
The expression of ${\cal A}_B^{(n^*-1)}$ is computed from
(\ref{Btau}), (\ref{tt}), and (\ref{tstart}).

There are two cases in Eq.~(\ref{tt}) corresponding
to the generic situation when $u$ is non-integer
and to the rather special case when $u$ is an integer.
They can be treated on the same footing by introducing the
`remainder':
\begin{eqnarray} 
r \equiv \left\{ 
\begin{array}{ll} 
\left\{ u\right\}\;,   & \mbox{if}\;\; u
\;\;\mbox{is non-integer}\;,\\ 
1\;, & \mbox{if}\;\; u
\;\;\mbox{is an integer}\;, 
\end{array} \right. \label{rt}
\end{eqnarray} 
where $\{ x\}$ denotes the fractional part of $x$ ($0< \{ x\} <1$).
Thus $n^*-1 = u-r$.

\begin{figure}[htbp]
\hspace*{0.8cm}
\begin{minipage}{5.0 in}\epsfxsize=5.0 in
\epsfbox{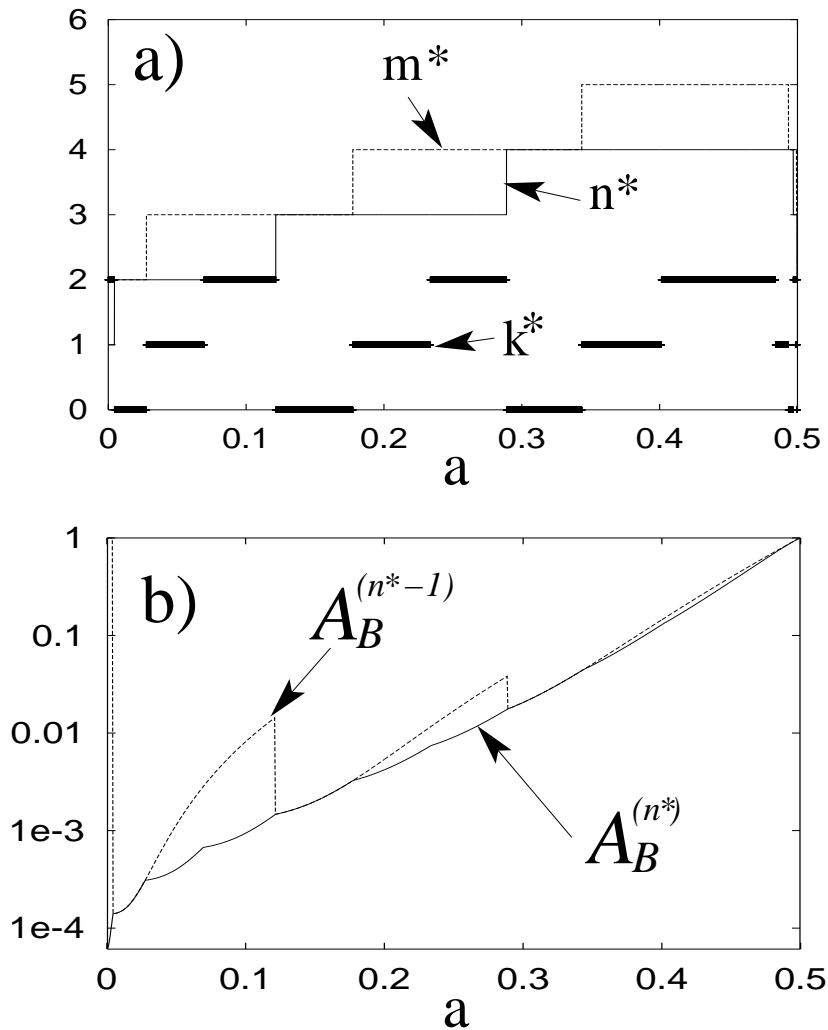}
\end{minipage}
\vspace*{0.5cm}
\caption{ 
The values  $m^*$, $n^*$ and $k^*$ at $\tau=3$
and $\sigma = 10^{-5}$ a), and the corresponding
steady state area 
${\cal A}_B^{(n^*)}$
of material $B$, together with 
${\cal A}_B^{(n^*-1)}$, 
 b). Note that the two areas coincide for $k^* = 0$, that
is when $m^*=n^*$.}\label{fig:compare}
\end{figure}

{From} Eq.~(\ref{tstart}) it follows that
$a^{u\tau} =(1+v/a)^{-1}$,  where
$v$ is given in (\ref{ve}).
Using this expression, and (\ref{Btau}), one obtains:
\begin{equation}
{\cal A}_B^{(n^*-1)} =
\left( 2^{u\tau}
\frac{2\sigma a \;C +
\left(
1-(2a)^{\tau}
\right)D}{
2\sigma a +\left(
1-a^{\tau}\right)(1-2a)}
-1\right) \frac{2\sigma}{ 
1-(2a)^{\tau}}\;, \label{steady}
\end{equation} 
where
\begin{eqnarray}
&& C=(2a)^{-r\tau} \; \frac{a^{\tau}(2^{\tau}-1)}
{1-a^{\tau}} + 2^{-r\tau} 
\frac{1-(2a)^{\tau}}
{1-a^{\tau}}\;,  \label{nice1} \\
&& D = (2a)^{-r\tau} \;a +
2^{-r\tau}(1-2a)\;. \label{nice2}
\end{eqnarray}
Unfortunately the coefficients $C$ and $D$ cannot be
simplified further, since they
contain, via $r$, the fractional-part function which usually
hinders analytical manipulations.

All these calculations refer to the area of $B$ occupied
after $n^*-1$ active-steps. 
When $m^*=
n^*$ (see Figures~\ref{fig:mns}) and \ref{fig:compare}a)) 
${\cal A}_B^{(n^*-1)}={\cal A}_B^{(n^*)}$ is the steady state area of $B$.
When $m^*=n^*+1$ 
(see Figures~\ref{fig:mns}) and \ref{fig:compare}a)) 
we have to substract from 
${\cal A}_B^{(n^*-1)}$ the `remainder'
$\sum_{k=1}^{k^*} m_{k}^{(n^*)}  
w_{k}^{(n^*)}$. 
This leads to a similar expression as (\ref{steady}),
but $C$ and $D$ are changed.
Thus for the steady state area we have
\begin{equation}
{\cal A}_B^{(n^*)} =
\left( 2^{u\tau}
\frac{2\sigma a \;C' +
\left(
1-(2a)^{\tau}
\right)D'}{
2\sigma a +\left(
1-a^{\tau}\right)(1-2a)}
-1\right) \frac{2\sigma}{ 
1-(2a)^{\tau}}\;, \label{steadyuj}
\end{equation} 
where $C'$ and $D'$ are shorthand notations for
\begin{eqnarray}
&& C'=2^{k^*-r\tau} \; \frac{1-(2a)^{\tau}}
{1-a^{\tau}}\left(1-a^{(1-r)\tau}\right) + (2a)^{(1-r)\tau}\;,
\label{niceuj1} \\
&& D' = (2a)^{k^*-r\tau} \;a +
2^{k^*-r\tau}(1-2a)\;. \label{niceuj2}
\end{eqnarray}
{From} these we recover (\ref{nice1}) 
and (\ref{nice2}) with $k^*=0$,
that is, for $m^*=n^*$, see figure 4.
This means that the steady states 
are described by (\ref{steadyuj}),
(\ref{niceuj1}), and (\ref{niceuj2}) in all cases. 
Note that the steady state area 
${\cal A}_B^{(n^*)}$
is a continuous function
of $a$ while ${\cal A}_B^{(n^*-1)}$ is not necessarily.
However, the derivative of ${\cal A}_B^{(n^*)}$ 
with respect to $a$ is discontinuous at all the points where
$k^*$ is discontinuous. 


In this section we have derived the dynamics of the areas covered by
materials $A$ and $B$.
Our investigations have been restricted, however, to the time instants
right after the reaction events.
In ~\ref{sec:app1} we include the 
whole area dynamics of the active 
baker map, i.e., we examine there what 
happens between subsequent reaction
events.

In the next section we make a connection
with the existing heuristical theory of active advection
and identify the geometric factor for the 
active baker map.


\section{Comparison with the heuristical theory: derivation
of the geometric factor}
\label{sec:compare}

The heuristical theory presented in the Introduction and
also in detail in
Ref.~\cite{pre, chaos} derives a recursion relation
for the area covered by material $B$ in arbitrary open flows 
just after
the $(n+1)$-st reaction as a function of
the area occupied by $B$ just after the $n$-th
reaction. 
This theory is valid
for purely hyperbolic systems.
In this section 
we show that in the
limit of sufficiently small $\sigma$ the phenomenological
expressions derived in \cite{pre,chaos} are recovered.

The heuristic formulas were derived with the
basic assumption that the material $B$ in the steady-state
is covering the unstable manifold of a chaotic set.
The interesting and novel reaction-kinetics is observed
in the parameter region (of $\sigma$, $\tau$, $D_0$ and
$\lambda$)
where the coverage is such that the hierarchical
structure of the (fractal) unstable manifold is preserved
in the system over several lengthscales. For example,
if  $\sigma$ is too large then the fractal filamental structure
of the unstable manifold is simply \emph{washed out} during
reactions, and thus the reaction-kinetics will be of
the common classical type of surface reactions \cite{classic}.
The effects of fractility of the manifold is seen when
$\sigma$ is small enough so that
the coverage of the filaments of the unstable manifold
preserves the filamental structures, in forms of
stripes of average width $\epsilon^*\ll 1$.
For the baker map this
means that the number of filaments in the steady state,
$2^{n^*\tau}$, is much greater than unity: 
$2^{n^*\tau} \gg 1$. {From} the  calculations above
follows that this regime is reached for `intermediate' $a$ values
at fixed $\tau$ and $\sigma \ll 1$. 
(If $a$ is close enough to either 0 or 0.5, $n^*$
is close to unity). However, the smaller $\sigma$
is the wider the `intermediate' $a$-range is where
fractal behaviour is important, see the expression of
$u$, Eq.~(\ref{tstart}). In particular for 
$\sigma = 10^{-5}$ the number of resolved levels 
is maximum ($n^* = 4$) in the interval $a \in (0.29,0.49)$,
see figure 4a). 
In the following we will assume that we are in this {\em scaling}
regime $\sigma\ll 1$. 

\begin{figure}[htbp]
\hspace*{0.0cm}
\begin{minipage}{5.8 in}\epsfxsize=5.8 in
\epsfbox{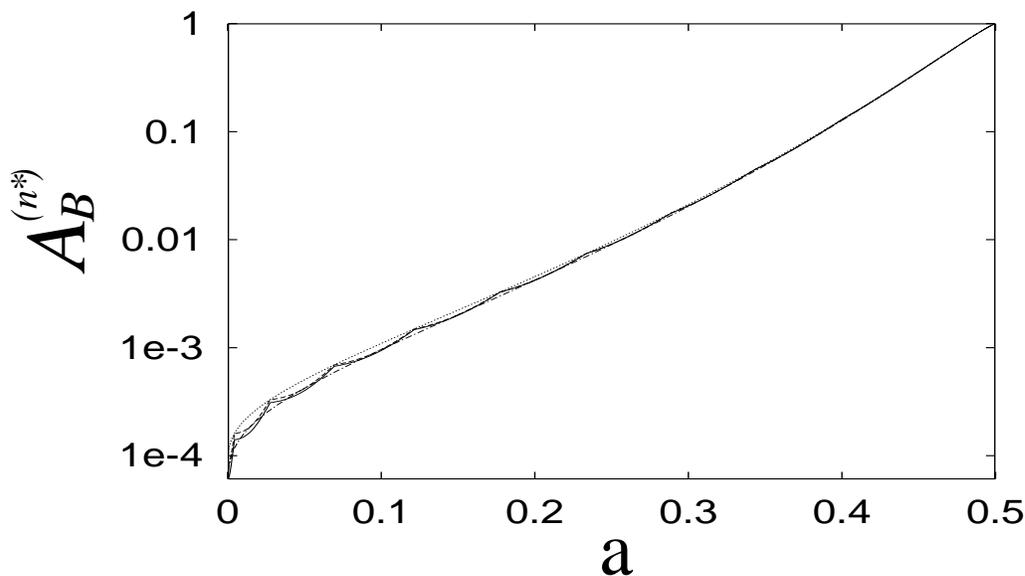}
\end{minipage}
\vspace*{0.5cm}
\caption{Comparing the approximate expression (\ref{steadyapp})
with the exact (\ref{steadyuj}).
The solid line is  Eq. \protect(\ref{steadyuj}), the dashed 
line is  Eq. \protect(\ref{steadyapp}). We have also
plotted an upper (using $g_u$, dotted line) and a lower bound 
(using $g_{\ell}$, dashed-dotted line) for 
Eq. \protect(\ref{steadyapp}).  
Here$\tau=3$ and $\sigma=10^{-5}$.
}\label{fig:compare1}
\end{figure}

First observe, that in the fraction multiplying
$2^{u}$ in Eq.~(\ref{steadyuj}) one can neglect
$\sigma$, since $\sigma \ll 1$.
Similarly, it is easy to show that $2^{u} \simeq
(v/a)^{-\ln{2}/\ln{a}}$ (according to (\ref{ve}), since
$v \gg 1$). It also follows that $2^{u} \gg 1 $
and the 1 can be neglected in the curly brackets of
(\ref{steadyuj}). Thus we end up with
\begin{equation}
{\cal A}_B^{(n^*)} \simeq
2^{u\tau}
\frac{2 \sigma D'}
{\left(
1-a^{\tau}\right)(1-2a)}
= e^{\kappa \tau}
\left(
\frac{\sigma g}{
e^{\kappa\tau/(2-D_0)}-1}
\right)^{2-D_0} \label{steadyapp}
\end{equation}
with $D_0$ and $\kappa$ given by Eqs.~(\ref{dimension}) and
(\ref{kappa}), and $D'$ by (\ref{niceuj2}). 
Here 
\begin{equation}
g=\frac{2a}{1-2a}\left(
\frac{D'}{a}\right)^{\frac{1}{2-D_0}}
=2\; \frac{\;\; (2D')^{\frac{1}{2-D_0}}\;}
{1-2a}\;, \label{ge}
\end{equation}
is a factor which mainly depends on the
characteristics of the chaotic baker dynamics.
Equation (\ref{steadyapp})
gives the expression of the steady state $B$-area just
after a reaction. This
coincides precisely with the heuristically derived
Eq.~(\ref{eq:map*})
(see also Eq.~(8) of Ref.~\cite{pre}), after employing the
relation $\kappa/(2-D_0)=\lambda$.
However, while in (\ref{eq:map*}) and in \cite{pre}
$g$ appears as a phenomenological constant, in the present
case of the active baker map, its expression can be given
explicitely.

\begin{figure}[htbp]
\hspace*{0.0cm}
\begin{minipage}{5.8 in}\epsfxsize=5.8 in
\epsfbox{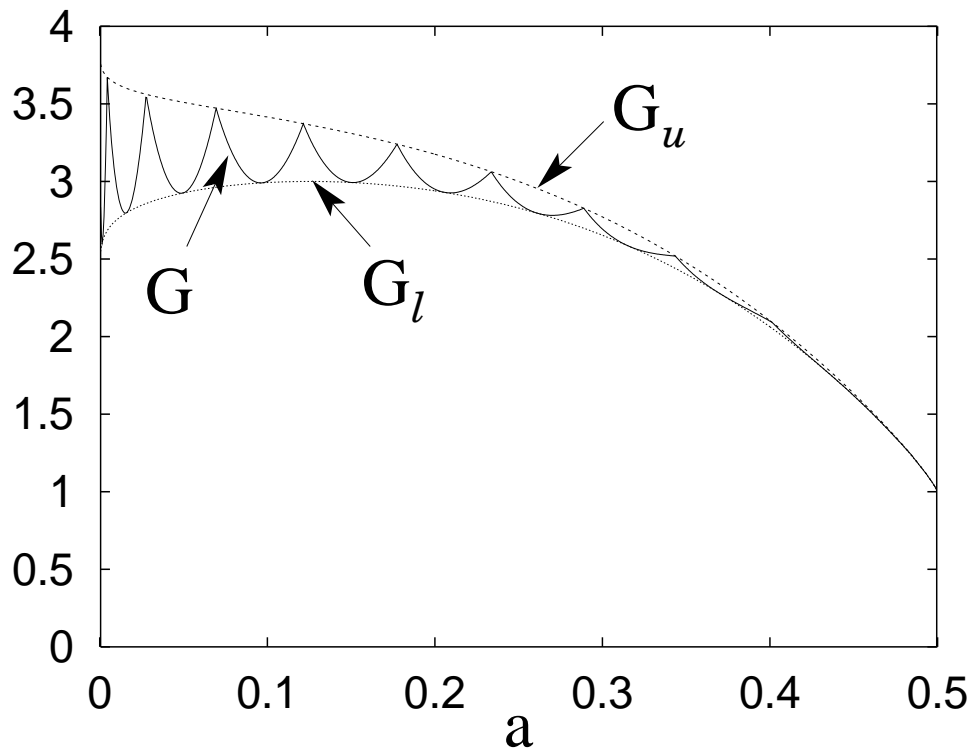}
\end{minipage}
\vspace*{0.5cm}
\caption{Comparing the geometrical coefficients $G$,
$G_{\ell}$  and $G_u$. Here
$\tau=3$ and $\sigma=10^{-5}$.
}\label{fig:compare2}
\end{figure}

Figure~\ref{fig:compare1} shows (on a linear-log scale) for 
comparison the $a$-dependence of the
steady state areas of $B$ for the exact expression of
${\cal A}^{(n^*)}_B$ from 
(\ref{steadyuj}) and the approximation in (\ref{steadyapp}).
Note that the two are virtually identical in the scaling regime
of $a$.

The explicit expression for $g$ is rather complex due to its dependence
on the integer-part function via $D'$, see Eq. (\ref{niceuj2}).
It can however be shown that $D'$ is a bounded variable. 

First,  $D' \leq 1-a$. For the geometric factor this leads to the
solely $a$-dependent upper bound, $g_u$:
\begin{equation}
g \leq
\frac{2a}{1-2a} \left( \frac{1-a}{a} \right)^{\frac{1}{2-D_0}}
\equiv g_u \label{bounds}
\end{equation}
Second, a lower bound can be derived for $g$, as follows. 
Let $F(x) =
a(2a)^{-x}+(1-2a)2^{-x}$, 
then $D'=F(r\tau-k^*)$. It can be shown that $0 \leq
r\tau-k^* <1$, and that the function $F$ has a single minimum in the $x \in
[0,1)$ interval, at: 
\begin{equation}
x_{\ell} = 1-\frac{1}{\ln{a}}\;
\ln{\left(
(1-2a)\;\frac{\ln{2}}{\ln{\frac{1}{2a}}}
\right)} \label{lowest}
\end{equation}
Thus, $D' \geq F(x_{\ell}) \equiv D'_{{\ell}}$ and therefore 
$g \geq g_{\ell}$,
where $g_{\ell}$ is given by (\ref{ge}) with $D'$ replaced by $D'_{\ell}$.

Note that for $a \to 1/2$, $g$ diverges since it is 
inversely proportional to $1-2a$. This divergence, is however cancelled 
in the expression for the area by the exponent $2-D_0$. In other
words, the quantity $g^{2-D_0}$ is well behaved ($1 \leq g^{2-D_0} \leq 4$).
Let us introduce therefore the quantity $G \equiv g^{2-D_0}$.
In figure 6 we show $G$, $G_{\ell}$ and $G_u$ as  functions
of $a$. Figure 5 shows the corresponding area dependence created with
both the upper bound (dotted line) and the lower bound (dashed-dotted line) .

Both bounds are independent on $\tau$ and $\sigma$. The two bounds
give a rather close approximation to the exact area expression, as shown in
figure 5.  This implies that the factor $g$ is weakly dependent
on the reaction range $\sigma$ and on the reaction time $\tau$, 
corroborating our claim that it is only {\em
geometrical}, i.e., it mainly depends on the parameters 
of the mixing dynamics
(in this case, on $a$).

It is also worth comparing the average width of the stripes of material
$B$ in the baker map to that obtained in the Intoduction
[see Eq.~(\ref{eq:MB0})]
and in Ref.~\cite{pre}.
The average stripe width in the steady state can be obtained as 
$\varepsilon^*={\cal A}_B^{(n^*)}/2^{(n^*-1)\tau+k^*}$, 
because there are
$2^{(n^*-1)\tau+k^*}$ stripes.
In the small $\sigma\ll 1$ range this gives
\begin{equation}
\varepsilon^*\approx\frac{2^{u\tau}}{2^{(n^*-1)\tau+k^*}}
\frac{2\sigma D'}{(1-a^{\tau})(1-2a)}=
\frac{g'\sigma}{1-a^{\tau}}\;,\label{epsstar}
\end{equation}
where $g'$ is given by
\begin{equation}
g'=2+\frac{2a^{k^*+1-r\tau}}{1-2a}\;.\label{geeps}
\end{equation}
Equation~(\ref{epsstar}) is exactly the same as (\ref{eq:MB0}) 
and the steady state of Eq.~(6)
of Ref.~\cite{pre} calculated after the reactions.
Note that geometrical factor (\ref{geeps}) differs from (\ref{ge}).
The reason is that in the heuristic theory of the Introduction and of
Ref.~\cite{pre} 
the fractal scaling relation 
${\cal A}_B\sim\epsilon^{2-D_0}$ was extended for the steady state
stripe width, that is, for $\varepsilon^*$.
For this value of $\epsilon$, we are at the edge of the
validity of the above mentioned scaling, since a crossover to a
two-dimensional behavior sets in.
This leads to different geometrical factors
for the areas and for the stripe widths.


\section{Another form of the baker map}
\label{sec:another}

In this section we slightly modify the baker map.
One reason to do so is to check whether the results obtained so far
are robust enough.
We show that only the geometrical factor is altered in this case,
the main results obtained in the Introduction and in
Refs.~\cite{prl,pre} are the
same. The other reason why another form of the baker map is introduced
is based on the observation that during the reactions there is material
hanging over the vertical edges 
(the edges parallel to the background flow) of the unit square.
The behaviour of this 
material is not described
by the baker map in the form of (\ref{bakerorig}).
Moreover, the cutting of these stripes hanging over the unit square is
quite unnatural in terms of 
hydrodynamics:
there is enough space for the growing of material across the background
flow, the growth is
usually not limited by such a sharp obstacle as the edge of the
unit square.

The modified baker map differs from (\ref{bakerorig})
in the location
of the fixed points around which the compression takes place.
It consists of two steps:
1) the lower and upper
half of the unit square is compressed in the $x$ direction
by a factor of $a < 1/2$ around
the fixed point in $(1/2-d/2(1-a),0)$ and 
$(1/2+d/2(1-a),1)$, respectively,
where $d$ is a parameter chosen appropriately, 
and then 2) the compressed
stripes are
stretched along the $y$ direction 
(around the same fixed points)
by a factor of $b=1/a> 2$.
In other words, after one baker-step we get 
two vertical (parallel to the
$y$ axis) stripes centered around $(1-d)/2$ and $(1+d)/2$.
Their centers are at a distance of $d$.
It can be chosen such that
the stripes do not reach the vertical edges of the
unit square, thus there remains enough space for the reactions.
An appropriate choice is, e.g.,  $d=1/2$.

The price for this modification is a more involved algebra.
In this case, after each baker-step there are empty areas 
(i.e., covered by
material $A$) inside the stripes of width $a$, 
into which the unit square
is mapped.
This means that the gaps are wider than in the former derivation,
and their calculation is not trivial.
In fact, it is more convenient to follow the distance of the centers
of the neighbouring stripes.
In a baker-step, all existing central distances are multiplied by $a$,
and a new level of the hierarchy is introduced.
This hierarchy changes only if there are gaps
disappearing.
Note that the stripe dynamics of the areas covered by material $B$ is
much simpler than in the former baker model: 
all stripes of material $B$ have equal widths.

Without going into details, we summarize the results obtained with this
modified baker model for the steady state areas.
First let us introduce
\begin{equation}
\sigma_{\mathrm{CR}}=\frac{d(1-2a)(1-a^{\tau})}{2(1-a)}>0\;.
\label{eq:sigcr}
\end{equation}
This is a critical value of the reaction distance for any choice of $a$
and $d$; if
$\sigma>\sigma_{\mathrm{CR}}$ then there is 
material $B$ hanging over the vertical edges of the unit square
in the steady state.
For $\sigma<\sigma_{\mathrm{CR}}$, which we assume in what follows,
we obtain a steady state
where a Cantor-like hierarchy is produced with $N$ different levels,
where $N$ is given by
\begin{eqnarray} 
N = \left\{ 
\begin{array}{ll} 
[z]+1\;,\;\;\; & \mbox{if}\;\; z
\;\;\mbox{is non-integer}\;,\\ 
z\;,\;\;\;  & \mbox{if}\;\; z
\;\;\mbox{is an integer}\;, 
\end{array} \right.\label{eq:Nsol}
\end{eqnarray} 
where
\begin{equation} 
z=
\frac{\ln{\left( \frac{\sigma_{CR}}{\sigma}\right)}}
{\ln{\left( \frac{1}{a}\right)}}\;. \label{zz} 
\end{equation} 
The steady state stripe width of material $B$ after the reactions is
\begin{equation}
\varepsilon^{*}=\frac{2\sigma}{1-a^{\tau}}+\frac{a^N d}{1-a}\;.
\label{eq:epssol}
\end{equation}

This is to be compared to the steady state stripe width (\ref{eq:MB0}),
using (\ref{dimension}) and (\ref{lambda}), which  hold for the
modified baker map as well.
{From} this equation and from (\ref{eq:epssol}) we
obtain
\begin{equation}
\overline{g}'=2+\frac{1-a^{\tau}}{1-a}\frac{a^N d}{\sigma}\;.
\label{eq:del}
\end{equation}
In case of this modified baker map we have also obtained a correction to
the $g=2$ value, which is due to the overlapping of stripes during the
reactions.
The geometrical factor depends mainly on parameter $a$ of the baker map,
as from (\ref{eq:sigcr}) and (\ref{eq:Nsol}) we obtain the bounds
\begin{equation}
2+\frac{2a}{1-2a}<\overline{g}'\leq 2+\frac{2}{1-2a}\;.
\label{b2}
\end{equation}

The total area covered by
material $B$ is
\begin{equation}
{\cal A}_B=2^N\varepsilon^{*}=
2^N\left(\frac{2\sigma}{1-a^{\tau}}+\frac{d a^N}{1-a}\right)\;.
\end{equation}
This is, again, in coincidence with that obtained in Eq.~(8) of
\cite{pre}, but the geometrical factor is different from
(\ref{eq:del}), it is
\begin{equation}
\overline{g}=\frac{1-a^{\tau}}{\sigma a^{\tau}}
\left(\frac{2^{N+\tau+1}\sigma a^{\tau}}%
{1-a^{\tau}}+
\frac{(2a)^{N+\tau}d}{1-a}\right)^{\frac{1}{2-D_0}}\;. \label{gbar}
\end{equation}
The deviation between $\overline{g}$ and $\overline{g}'$ 
is again the result of 
the extension of the fractal scaling
to the range of $\varepsilon^*$, to the steady state stripe width.

\begin{figure}[htbp]
\hspace*{0.2cm}
\epsfxsize=5.8 in 
\epsfbox{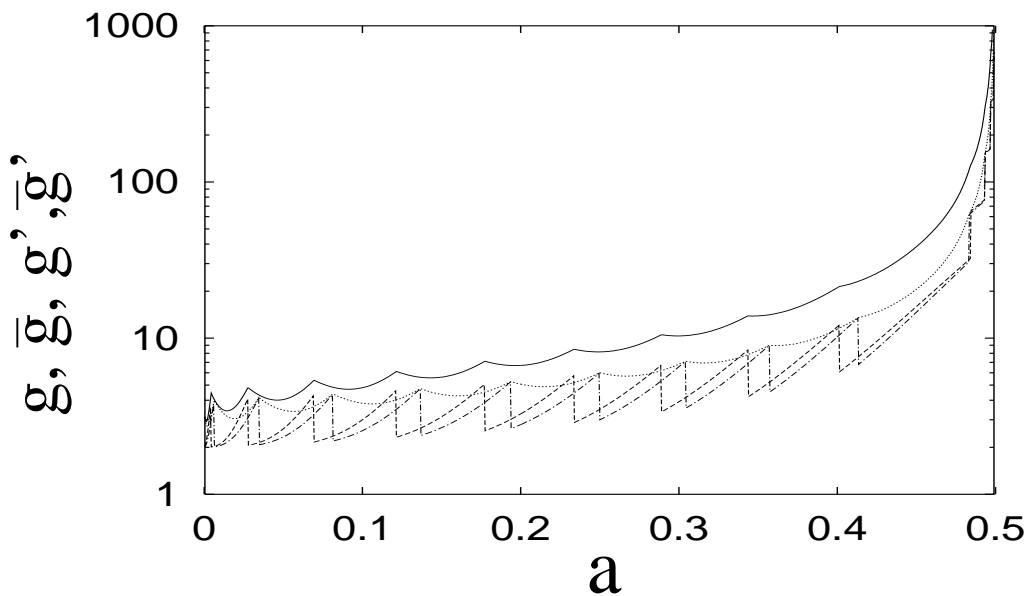}
\vspace*{0.0cm}
\caption{Compare the geometric factors from Eqs. (\ref{ge})
(solid line),  (\ref{geeps}) (dashed line), and Eqs. (\ref{gbar})
(dotted line), (\ref{eq:del}) (dashed-dotted line).
$\tau=3, \sigma=10^{-5}$, $d=1/2$.
}\label{fig:thegees}
\end{figure}

In Figure \ref{fig:thegees} we compare the 
four geometrical factors (\ref{ge}) (solid line)
and (\ref{gbar}) (dotted line), and 
(\ref{geeps}) (dashed line), (\ref{eq:del}) (dashed-dotted line)
as function of
$a$. Although the average trend of the curves is the same,
there are large deviations due to the fact that $g'$ and 
$\overline{g}'$ are discontinuous functions. The
deviations, however are less pronounced in the scaling region
where the fractal structures of the maps are better resolved.
This leads us to the conclusion that the geometric factor
strongly depends on the particular realizations and relative
positions of the filaments even when these realizations share
the same fractal and dynamical properties ($D_0$, $\lambda$).


\section{Discussions and outlook}
\label{sec:discussion}

 In this paper we have investigated the effects of a fractal 
geometric catalyst supplied e.g.\ by the Langrangian dynamics of an
underlying flow on the reaction kinetics of autocatalytic
reactants transported in this flow. The reactive particles
were assumed to simply follow the local velocity of the
fluid element at their position and act as simple passive
tracers that do not modify the properties of the flow. 
For mixing dynamics we have considered the simplest
possible dynamics,namely two variants of the baker
map,
in order to be able to
generate an analytic
description of the reaction process. 
Even though the baker map is the simplest map with
chaotic mixing, when reactions are superimposed on the 
transport dynamics, the analytical calculations become 
involved since the reaction range $\sigma$ introduces a 
{\em finite nonscaling
cutoff} which breaks the strict selfsimilarity. 
In the limit of slow reactions
one is able to recover from these exact formulas 
the heuristic expressions derived for general flows, 
thus validating the latter approach,
and showing that indeed, the fractal backbone does enhance
the productivity of reactions, and acts as a fractal catalyst.

\begin{figure}[htbp]
\hspace*{0.2cm}
\epsfxsize=5.8 in 
\epsfbox{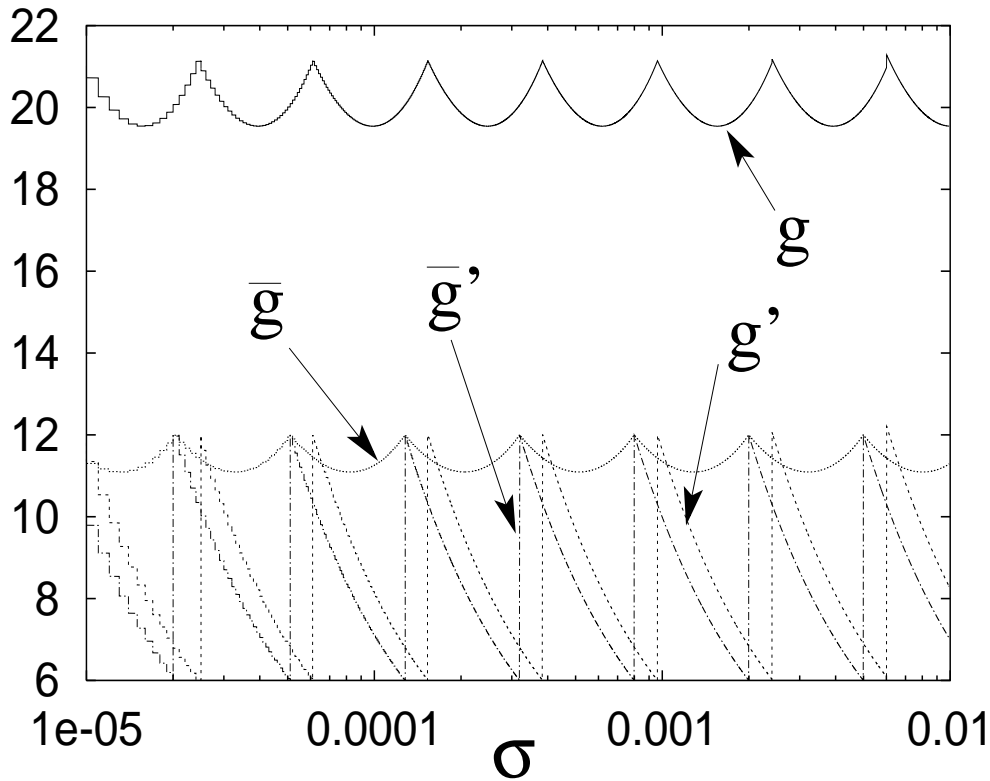}
\vspace*{0.0cm}
\caption{The dependence of the geometric factors 
$g$, $\overline g$, $g'$, $\overline{g}'$ on the reaction
range, $\sigma$.
 Here $a=0.4$, $\tau=3$, and $d=1/2$.
}\label{fig:thegeess}
\end{figure}

A crucial point in the consistency between the heuristic and exact 
approach is played by the so-called geometrical factor $g$
(or its variants $g', \bar{g}, \bar{g'}$).  By definition, it should
be a parameter reflecting mainly the geometrical
properties of the mixing dynamics (of the hydrodynamical flow). In the
heuristic theory it is postulated to be independent of
the chemistry parameters $\sigma$ and $\tau$.
The consistency is achieved by showing that it is only		
{\em weakly} depending on these parameters, in the scaling limit at least.
This property follows from the close bounds
(\ref{bounds}) and (\ref{b2}) on the geometrical factor(s)		
depending on $a$ only.
A direct plot of the geometrical factors as a function of the reaction range
$\sigma$ (figure 8) also supports this property and suggests a weak, 
approximately periodic dependence on the 
logarithm of the reaction range, over several orders of magnitude.
Note, that the actual value of the geometrical factor strongly depends on the
two independent global chaos parameters ($\lambda$, $D_0$) which 
determine the chemical kinematics (see Introduction), but might also
depend on other aspects of the chaotic dynamics (as e.g., on the parameter
$d$ of the modified baker map). 

Various improvements and extensions to the present theory
are possible: 1) one can consider \cite{giovanni,NZ1} 
the more realistic situation
of non-simultaneous reactions, i.e., the particles have a random
phase dispersed among them. We expect that this only leads
to a more rough, fluctuating $A-B$ interface along the 
manifolds, however these fluctuations are bounded due to 
the fact that only  correlations of finite length can 
develop along the interface during the finite 
time lag $\tau$, and thus it will renormalize
the front velocity;
2) the interacting particles (which can be
for example phytoplankton \cite{pnas}) may locally modify the
advection dynamics
due to inertia and finite size effects. In this case
the phase space of the problem increases in dimensionality
by the number of velocity components of the particles since the
latter no longer is the same as the fluid elements' velocity;
3) and at last but not the least, various other chemical
reactions should be monitored, besides the autocatalytic ones
presented here.

Finally, it is worth connecting
the baker map results with the general description given in 
the Introduction. To this order, we mention that in a map
of linear extension $L$, the reaction range is $\sigma L$. Analogously,
the dimensional reaction time is 
$\tau T$, where $T$ is  the temporal period over which
the baker map is acting. 
The reaction velocity $v_r$ is then 
\begin{equation}
v_r = \frac{\sigma L}{\tau T}.
\label{vr}
\end{equation}
Taking the continuum limit of the reaction (see Introduction)
means that both the reaction range and time are short on the corresponding
macroscopic scales. Since $L$ is interpreted as a hydrodynamical
length, this implies $\sigma \ll 1$ as used in the baker map. On the 
other hand, since $\tau$ was taken to be of order unity,
$\tau T$ can only be short if the duration of the map is much
{\em shorter than the characteristic hydrodynamical time}.
Even without further specifying the choice of $T$, we can derive a useful
condition  in terms of a dimensionless number.
 
In the case of reactions in fluid flows with chaotic mixing
properties we have three distinct levels of time-continuous dynamics
built on top of each other: 
\begin{itemize} 
\item the Eulerian description of the 
hydrodynamics, characterized by one main parameter, the Reynolds
number
\begin{equation}
Re= \frac{UL}{\nu}\;,
\end{equation}
where $U$ is a typical velocity, $L$ is the typical length scale
and $\nu$ is the kinematic viscosity;
\item the Langrangian description of the dynamics, characterized
by the fractal dimension of the `transport manifolds' (unstable manifolds), 
and by the
contraction dynamics on them,
\begin{equation}
D_0=2-\frac{\kappa}{\lambda}\;,\;\;\;\lambda=\lambda (Re)\;,\;\;\;
\kappa=\kappa (Re)\;;
\end{equation}
\item the reaction kinetics, described by a `reaction' P\`eclet number,
\begin{equation}
(Pe_r)^{-1} = \frac{v_r}{\lambda L}\;,\;\;\;Pe_r=Pe_r(\lambda (Re)).
\label{Pe}
\end{equation}
\end{itemize}
Reaction equation (\ref{diffeq}) can be made dimensionless
if the area is measured in units of $L^2$, and the time in units
of the characteristic lifetime of (the Lagrangian) chaos, 
$1/\kappa=1/(\lambda(2-D_0))$:
\begin{equation}
\dot{{\cal A}}_B=-{\cal A}_B+
g \; \frac{1}{Pe_r}
\left({\cal A}_B
\right)
^{-\beta}. \label{diffeqd}
\end{equation}
The active baker map's P\`eclet number is
\begin{equation}
(Pe_r)^{-1} = 
\frac{\sigma }{\tau T \lambda }=
\frac{\sigma}{\tau \ln{(1/a)}}.
\end{equation}
Here we have used that the dimensional Lyapunov exponent of the map is
$- \ln{a} /T$. With our typical choice of $\sigma = 10^{-5}$, 
$\tau=1,2,3$, and $a$ on the order of $1/4$, the reaction
P\`eclet number is much larger then unity.

It is worth giving the definition of 
the traditional P\`eclet number \cite{Liggett}
which measures the importance of {\em diffusion} relative to the flow:
\begin{equation}
(Pe)^{-1} \equiv \frac{D}{UL}
=\frac{v_d}{U}
\label{Pee}
\end{equation}
where $v_d=D/L$ is the charactersitic diffusion velocity. Large $Pe$
implies weak diffusion.
As suggested by the analogy between
the traditional and the reaction   P\`eclet number, the reaction 
range plays a similar role as a diffusion length and the reaction
velocity $v_r$ as $D/L$.
Furthermore, if molecular diffusion is present
but not very strong,
it only renormalizes 
the reaction front 
velocity and all the formulas shown here are valid with $v_r$ replaced
by an effective reaction front velocity \cite{chaos}.

Note that the direct reaction analog of the  P\`eclet number would be 
$v_r/U$. It is worth pointing out however that instead we have 
\begin{equation}
(Pe_r)^{-1} = 
\frac{v_r}{U}
\frac{U}{\lambda L} = \frac{v_r}{U}
 \frac{\mathrm{Lyapunov \;time}}
{\mathrm{hydrodynamical \;time}}
\end{equation}
which is rather a \emph{Lagrangian dynamics} based quantity.
The difference between $Pe_r$ and its hydrodynamical analog,
${v_r}/{U}$, might be pronounced if
the hydrodynamical
and Lyapunov times are not on the same order of magnitude.

The  
limit of slow reactions implies $v_r \ll \lambda L$,
i.e., $Pe_r \gg 1$. It is in this limit where pronounced
fractal product distributions are expected.
The steady state product area ${\cal A}_B^*$ 
given by Eq.~(\ref{contarea}) 
is by a factor $(Pe_r)^{D_0-1}\gg 1$
larger than it would be in a nonchaotic flow with $D_0=1$.
Thus we find that a large value of the
reaction  P\`eclet number introduced by Eq.(\ref{Pe}) provides 
a 
general, neccessary condition for the existence of
filamantal product distributions observable over several orders 
of magnitude in reactions taking place in chaotic flows.


\ack

The authors would like to express special thanks to   
I. Scheuring for a very helpful input and discussions.
Z.T. was supported by the Department of
Energy, under contract W-7405-ENG-36. G. K. and T. T.
acknowledge support of the Hungarian Science Foundation under grant
numbers T032423, F 029637, T 032423 .
G.K. is also indebted to the Bolyai grant and to 
FKFP 0308/2000 for financial
support.
Additional support was provided by the MTA-OTKA/NSF-INT526 grant and
the US-Hungarian Joint Grant (Project No. 501).

\appendix

\section{Area-dynamics in the active baker map}
\label{sec:app1}

In Sec.~\ref{sec:active} we described the hierarchy
of the stripes formed in the active baker map just after
each reaction, calculated the total area
occupied by the materials $A$ and $B$ as a function
of the number of \emph{active-steps}, and proved that a steady
state sets in after a given  number of active-steps,
$m^*$. 
In this appendix we present
the full temporal dynamics of the areas occupied by $A$
and $B$ (not only as a function of the active-steps).

 In order to account for the `intermediate' steps as well,
we will extend a little bit the notation introduced in
Sec.~\ref{sec:active}.
First we have to observe that once a reaction
has been completed and new passive baker-steps are taken, the
existing hierarchy at the end of the last reaction is doubled 
and mapped on a  level one step lower in the hierarchy
after a passive baker-step, 
while at the top of
the hierarchy
there are reaction-free gaps appearing. If $\rho$ denotes
the number of reaction-free baker-steps taken just after the last
reaction, the top gap of width $1-2a$  appears at
$\rho=1$, the additional two gaps of width $a(1-2a)$ 
appear at $\rho=2$, etc., until at $\rho=\tau$ the
reaction-free tree becomes completed. 
All the
rest of the gaps coming from earlier reactions are
lying under this newly created `baker tree' of length $\tau$.
Next a reaction occurs and the whole process repeats itself.

\begin{figure}[htbp]
\hspace*{0.2cm}
\epsfxsize=5.8 in 
\epsfbox{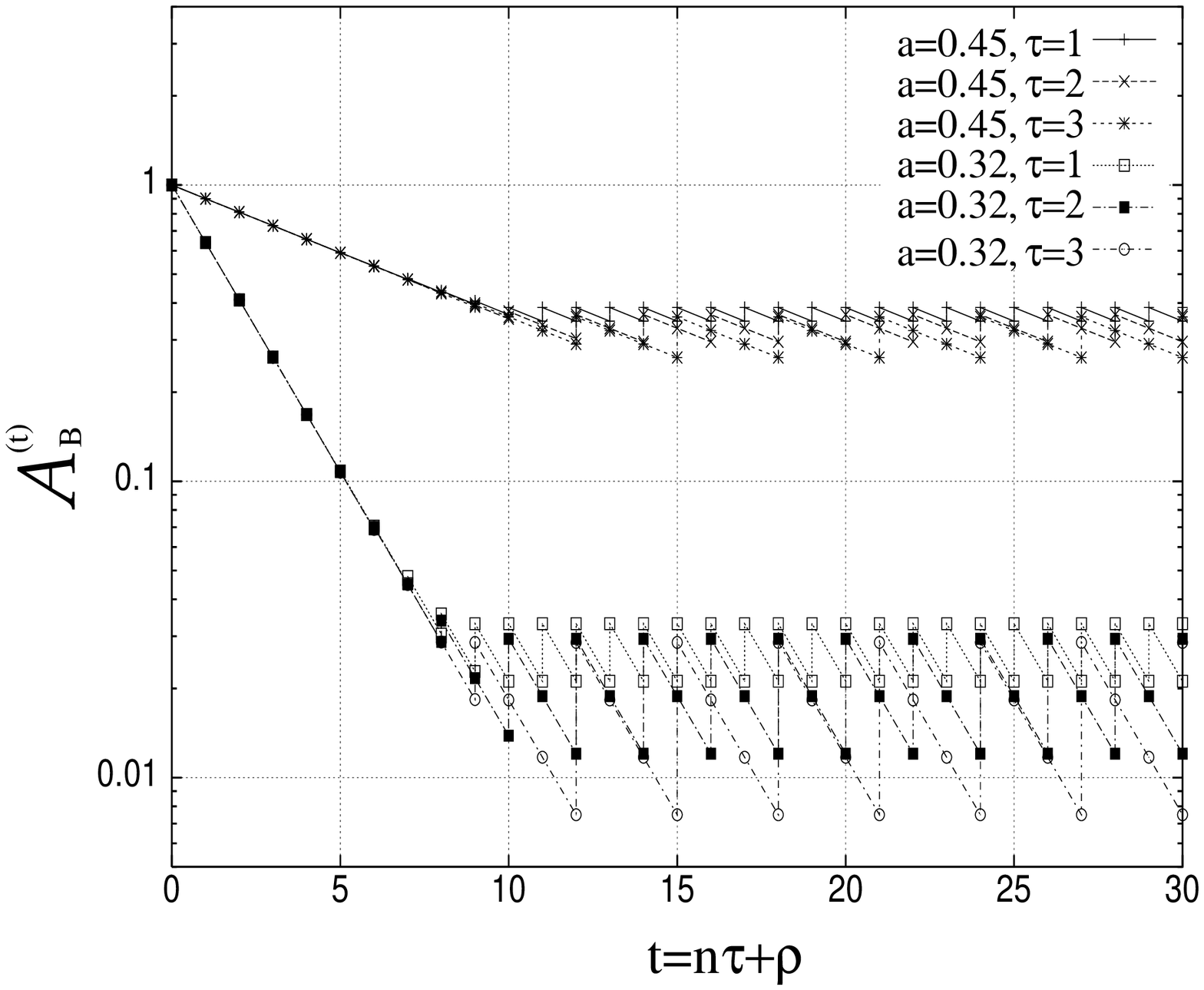}
\vspace*{0.0cm}
\caption{The
evolution of the area of material $B$ shown 
on a log-linear scale for three
values of $\tau$ and two values of $a$. In all 
cases $\sigma=10^{-5}$.
}\label{fig:areas}
\end{figure}

Let us denote the widths of the gaps coming from $\rho$
baker-steps by $w^{(0,\rho)}_\ell$, $\ell\in \{ 1,2,..,\rho\}$.
We have $w^{(0,\rho)}_{\ell} = a^{\ell-1}(1-2a)$ and the corresponding
multiplicities are $m^{(0,\rho)}_{\ell} = 2^{\ell-1}$. 
We shall denote the
widths and multiplicities of the gaps resulted
from previous $n$ reactions  and the additional $\rho$ passive
baker-steps by $w^{(j,\rho)}_k$, and  $m^{(j,\rho)}_k$, 
respectively, where
$k =1,2,..,\tau$ and $j =1,2,..,n$
(the sub-trees coming from previous reactions are all completed). 
To connect with the notation from Sec.~\ref{sec:active}, observe
that $w^{(j,0)}_k = w^{(j)}_k$.
We have $w^{(j,\rho)}_k = a^{\rho} w^{(j)}_k$, and the corresponding
multiplicity of $m^{(j,\rho)}_k = 2^{\rho}m^{(j)}_k$,
$k =1,2,..,\tau$ and $j =1,2,..,n$.

When the $(n+1)$-st reaction
occurs we obtain the widths $w^{(j)}_k$, where $k =1,2,..,\tau$
and $j =1,2,..,n+1$, with
$w^{(j+1)}_k =
w^{(j,\tau)}_k-2\sigma$, $m^{(j+1)}_k =
m^{(j,\tau)}_k$ (and it also leads to 
Eq.~(\ref{tausol})). 
The area of material $A$
($B$) after $n$ reactions and $\rho$ reaction-free
baker-steps will be denoted
by ${\cal A}^{(n,\rho)}_A$ (${\cal A}^{(n,\rho)}_B$). By definition,
${\cal A}^{(n,0)}_A \equiv {\cal A}^{(n)}_A$
(${\cal A}^{(n,0)}_B \equiv {\cal A}^{(n)}_B$).
The areas can be calculated
as follows:
\begin{equation}
{\cal A}^{(n,\rho)}_A =
\sum_{\ell=1}^{\rho} m^{(0,\rho)}_{\ell}
w^{(0,\rho)}_{\ell} +
\sum_{j=1}^{n}\sum_{k=1}^{\tau}
m^{(j,\rho)}_k w^{(j,\rho)}_k = 1-(2a)^{\rho}+(2a)^{\rho}
{\cal A}^{(n)}_A \label{tort}
\end{equation}
where ${\cal A}^{(n)}_A$ is given by (\ref{Atau}).
The above expression is valid in the regime where
no gap fillings have occured yet. In terms of the
$B$-material, using (\ref{kappa}), (\ref{tort}) is  equivalent to
\begin{equation}
{\cal A}^{(n,\rho)}_B =
(2a)^{\rho}
{\cal A}^{(n)}_B = e^{-\kappa \rho} {\cal A}^{(n)}_B \label{tortB}
\end{equation}
just as expected, since between two reactions
there is an exponentially emptying dynamics.
The total time is 
$t=n \tau +\rho$, $\rho =1,2,..,\tau$. When $\rho=\tau$,
$t=(n+1)\tau$ and a reaction occurs instantaneously.

Figure~\ref{fig:areas} shows the
$t$-dependence of the area covered by the material 
$B$ for three $\tau$ values, for given 
$\sigma$ ($\sigma=10^{-5}$)
on a log-linear scale 
at two $a$ values, $a=0.45$ and $a=0.32$.
Notice the fact that the actual time evolution of the
areas follow closely 
the simple exponential emptying dynamics. Certainly,
between two consecutive reactions, the emptying
\emph{is} exponential, the deviation from the
exponential in the overall run of the curves
occurs in jumps at reaction events. 

In order to
observe a longer temporal evolution before
stationarity sets in, one needs to choose a very
small $\sigma$. 
On the other hand, this means
that the overall fattening during a reaction is very weak,
and it is in fact a `perturbation' at early times.
It becomes `visible' just before the steady state.


\end{document}